\documentclass[a4paper,nodate]{sab} 

\usepackage{graphicx}
\usepackage[backend=biber, 
            natbib=true, 
            sortcites, 
            sorting=ynt, 
            doi=false, 
            url=false, 
            eprint=false, 
            style=apa
            ]{biblatex}
\usepackage{csquotes}
\usepackage{txfonts} 
\usepackage{hyperref}
\usepackage{url}
\usepackage[T1]{fontenc} 
\usepackage{t1enc}
\usepackage{xcolor} 

\newcommand{\gaiaeso}{\textit{Gaia}-ESO}
\newcommand{\gaia}{\textit{Gaia}}
\newcommand{\feh}{[Fe/H]}
\newcommand{\rbirth}{$\langle R_{\rm b} \rangle$}
\newcommand{\rgui}{$\langle R_{\rm g} \rangle$}

\renewcommand\ackname{Acknowledgements.} 
\renewcommand\andname{\&}
\renewcommand\keywordname{Keywords.} 
\renewcommand\figureptname{Figure}
\renewcommand\tableptname{Table}

\renewcommand\noteaddname{Note added to proofs.}
\renewcommand\andname{\&}

\idline{Boletim da Sociedade Astron\^omica Brasileira, {\bf 31}, no. 1, XX-XX}{1}

\begin{document} 

\title{The Milky Way in motion: gauging stellar trajectories that shape the Galactic thin disc} 

\author{
    M.~L.~L.~Dantas\inst{1,2}
    \and
    R.~Smiljanic\inst{3}
    \and
    R.~S.~de Souza\inst{4,5,6}
    \and
    P.~B.~Tissera\inst{1,2}
    \and
    L.~Magrini\inst{7}
    }

\institute{
    Instituto de Astrofísica, Pontificia Universidad Católica de Chile, Av. Vicuña Mackenna 4860, Santiago, Chile
    \and
    Centro de Astro-Ingeniería, Pontificia Universidad Católica de Chile, Av. Vicuña Mackenna 4860, Santiago, Chile\\
    \email{mlldantas@protonmail.com}
    \and
    Nicolaus Copernicus Astronomical Center, Polish Academy of Sciences, ul. Bartycka 18, 00-716, Warsaw, Poland
    \and
    Centre for Astrophysics Research, University of Hertfordshire, College Lane, Hatfield, AL10~9AB, UK
    \and 
    Instituto de Astronomia, Geofísica e Ciências Atmosféricas, Universidade de São Paulo, Rua do Matão 1226, 05508-090, São Paulo, Brazil
    \and 
    Department of Physics \& Astronomy, University of North Carolina at Chapel Hill, NC 27599-3255, USA
    \and
    INAF -- Osservatorio Astrofisico di Arcetri, Largo E. Fermi 5, 50125 Firenze, Italy
    }

\date{Received } 

\Abstract{As stars traverse the Milky Way, their orbits evolve through perturbations that alter their orbital radii. These changes arise from two mechanisms: \emph{churning}, which modifies angular momentum, and \emph{blurring}, which induces eccentric orbits without major angular momentum change. To assess whether churning or blurring dominates the dynamical evolution of \gaiaeso\ stars, we refine Galactic chemical-evolution models by constructing finer grids that span a wider age range. Using a generalised additive model (GAM), we estimate stellar birth radii beyond the limits of binned metallicity models and compare them with dynamical parameters derived from \gaia\ parallaxes and proper motions, and \textsc{Galpy}. Our metallicity-stratified sample, grouped through hierarchical clustering of 21 chemical abundances, reveals clear migratory signatures: metal-rich stars formed in the inner disc preferentially move outwards, while more metal-poor stars formed at larger radii tend to migrate inwards. About 75\% of stars show signs of churning, while the remainder are largely undisturbed or shaped by blurring. These patterns vary among chemical groups, likely reflecting interactions with the Galactic bar and spiral arms.}
{À medida que as estrelas percorrem a Via Láctea, suas órbitas evoluem devido a perturbações que alteram seus raios orbitais. Essas mudanças decorrem de dois mecanismos: o \emph{churning}, que modifica o momento angular, e o \emph{blurring}, que induz órbitas excêntricas sem grande alteração de momento angular. Para avaliar se \emph{churning} ou \emph{blurring} dominam a evolução dinâmica das estrelas do levantamento \gaiaeso, refinamos modelos de evolução química Galáctica ao construir grades mais finas que abrangem uma faixa etária mais ampla. Utilizando um modelo aditivo generalizado (GAM), estimamos os raios de nascimento estelares para além das limitações dos modelos binados em metalicidade e os comparamos com parâmetros dinâmicos derivados das paralaxes e movimentos próprios do \gaia\ e do código \textsc{Galpy}. Nossa amostra estratificada em metalicidade, agrupada via \emph{clustering} hierárquico de 21 abundâncias químicas, revela assinaturas migratórias claras: estrelas ricas em metais formadas no disco interno migram preferencialmente para fora, enquanto estrelas mais pobres em metais formadas em raios maiores tendem a migrar para dentro. Cerca de 75\% das estrelas apresentam indícios de \emph{churning}, enquanto o restante permanece amplamente inalterado ou é moldado principalmente pelo \emph{blurring}. Esses padrões variam entre os grupos químicos, provavelmente refletindo interações com a barra Galáctica e os braços espirais.}



\maketitle 

\section{Introduction}
\label{sec:intro}

The Milky Way (MW) is a barred spiral galaxy, most likely of SBbc classification \citep{KormendyBender2019}, composed of several interconnected stellar components, each preserving distinct clues to its formation history. Its thin and thick discs dominate the stellar mass in the Solar neighbourhood, with the thin disc hosting younger, more metal-rich, and kinematically cooler populations, while the thick disc is older, more metal-poor, and dynamically hotter \citep{Gilmore1983, Bensby2014, Bland-HawthornGerhard2016}. The central regions are structured around a boxy/peanut-shaped barred bulge, shaping the stellar orbits in the inner Galaxy \citep{Wegg2013, Bland-HawthornGerhard2016, Barbuy2018, Zoccali2026}. Surrounding these disc components lies the stellar halo, a diffuse and dark-matter- and accretion-dominated structure with wide metallicity and kinematic distributions \citep{Helmi2020}.

Although the MW disc is often described in terms of distinct structural components, stars do not remain fixed at the radii where they formed. Instead, their angular momenta can evolve over Gyr timescales through interactions with Galactic non-axisymmetric features such as spiral arms, the central bar, and transient perturbations. These processes, collectively known as radial migration or churning, change a star’s guiding radius by altering its angular momentum while tending to reduce its orbital eccentricity. This contrasts with blurring, in which stars oscillate around the same guiding radius through epicyclic motions, increasing their radial excursions without experiencing a net change in angular momentum. As a consequence, the present-day chemo-dynamic distributions of disc stars are the result of both churning and blurring, producing extensive dynamical reshuffling that complicates any attempt to infer the Galaxy’s evolutionary history from current positions alone \citep[e.g.][]{SellwoodBinney2002, Schonrich2009a, Frankel2020}.

Quantifying the extent of stellar mixing across the Galaxy is challenging because not all components preserve informative chemo-dynamic signatures. The bulge is dominated by complex bar-driven orbital families, rapid early star formation, and overlapping chemical sequences, which severely hinder any attempt to recover birth radii or past migration \citep[e.g.][]{Barbuy2018, RojasArriagada2020}. The thick disc exhibits large age uncertainties, broader metallicity distributions, and substantial dynamical heating, all of which blur the imprint of past radial motions \citep{Minchev2013, Buck2020}. The stellar halo, in turn, is thought to arise predominantly from accretion and merger events \citep{Helmi2020}, though evidence for an \emph{in situ} component---including stars possibly heated from the early disc---indicates that part of the halo may be more closely linked to the Galaxy’s internal evolution than previously assumed \citep[e.g.][]{Bonaca2017}. By contrast, the thin disc retains comparatively tight age--metallicity--kinematic relations and clearer chemical gradients, making it the most suitable environment in which to quantify radial migration and its contribution to disc evolution.

In this context, our recent work \citet{Dantas2025a} developed a data-driven framework to estimate stellar birth radii using chemo-dynamic information from the \gaiaeso\ survey \citep[][]{Gilmore2022, Randich2022}. By combining the widely spaced bins of the chemical-enrichment models from \citet{Magrini2009} with a generalised additive model (GAM), we directly inferred the birth locations of our thin disc stars and compared them with their current guiding radii, allowing us to quantify inward and outward migrators and to characterise the degree of radial mixing. In this proceedings contribution, we briefly summarise the main results of \citet{Dantas2025a} and present additional visualisations that further highlight the underlying chemo-dynamic trends.

\section{Data and methodology}
\label{sec:data}

Our analysis is based on the high-resolution \gaiaeso\ survey sample described in \citet{Dantas2023, Dantas2025a}, comprising 1188 thin disc FGK stars observed with UVES ($R \approx 47\,000$) and processed via the homogenised iDR6 pipeline \citep{Sacco2014, Smiljanic2014, Hourihane2023, Worley2024}. For each star, we adopted atmospheric parameters and detailed chemical abundances (21 species of 18 elements), complemented by ages from \textsc{Unidam} \citep{Mints2017, Mints2018} via PARSEC isochrone fitting \citep[][]{Bressan2012}, and orbital parameters computed with \textsc{Galpy} \citep{Bovy2015} using \gaia\ EDR3 astrometry \citep[][]{GaiaEDR3} and the \citet{McMillan2017} potential. Following \citet{BoessoRochaPinto2018} and the full description in \citet{Dantas2023}, we applied a hierarchical clustering (HC) analysis to these abundances, grouping stars into chemically similar stellar populations. The data selection and treatment is fully described in \citet{Dantas2023}.

To estimate the birth radii of these stars, we constructed a GAM using the radial metallicity profiles of \citet{Magrini2009}, which provide Galactocentric radius ($R$) as a function of \feh\ gradients and formation epoch ($t$). The GAM extends these sparsely binned theoretical profiles into a smooth, continuous surface $R(\rm{\feh}, t)$, enabling us to infer the most likely birth radius ($R_{\rm b}$) for any star given its metallicity and age. We used the \texttt{mgcv} package in the \textsc{r} environment to implement the GAM with smoothing splines, including a two-dimensional interaction term to capture non-linear couplings between \feh\ and formation time. The method is described and discussed in detail in \citet{Dantas2025a}.

Each star’s $R_{\rm b}$ was then derived by applying the GAM to \feh\ and its formation epoch, defined as
\mbox{$t = 13.8~\mathrm{Gyr} - t_{\star}$}, where $t_{\star}$ is the median stellar age from \textsc{Unidam}. Uncertainties were evaluated using a bootstrap procedure in which stellar parameters (\feh, $t$) were resampled 1000 times following \citet{Dantas2023}. This provides posterior distributions, quantiles, and robust standard errors for $R_{\rm b}$. Throughout, \rbirth\ denotes the median of each star’s bootstrapped \rbirth\ distribution, and we compare it with the current guiding radius (\rgui) to identify inward-, outward-, and non-migrating populations across the chemically defined stellar groups.

\section{Results, discussion, and final remarks}
\label{sec:results}

\begin{figure*}
    \centering
    \includegraphics[width=\linewidth, trim={0 0.6cm 0 0.5cm}, clip]{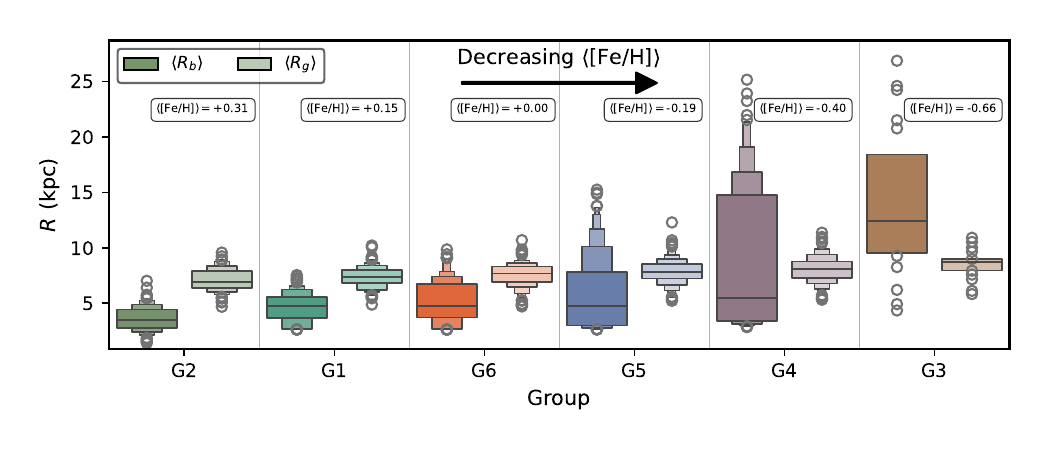}
    \caption{Letter-value plots showing the distribution of inferred birth radii (\rbirth, in dark shades, left distribution of each group) and present-day guiding radii (\rgui, in light shades, right distribution of each group) across the six chemically identified groups.}
    \label{fig:boxenplot}
\end{figure*}

\begin{figure}
    \centering
    \includegraphics[width=\linewidth, trim={0 0.5cm 0 0}, clip]{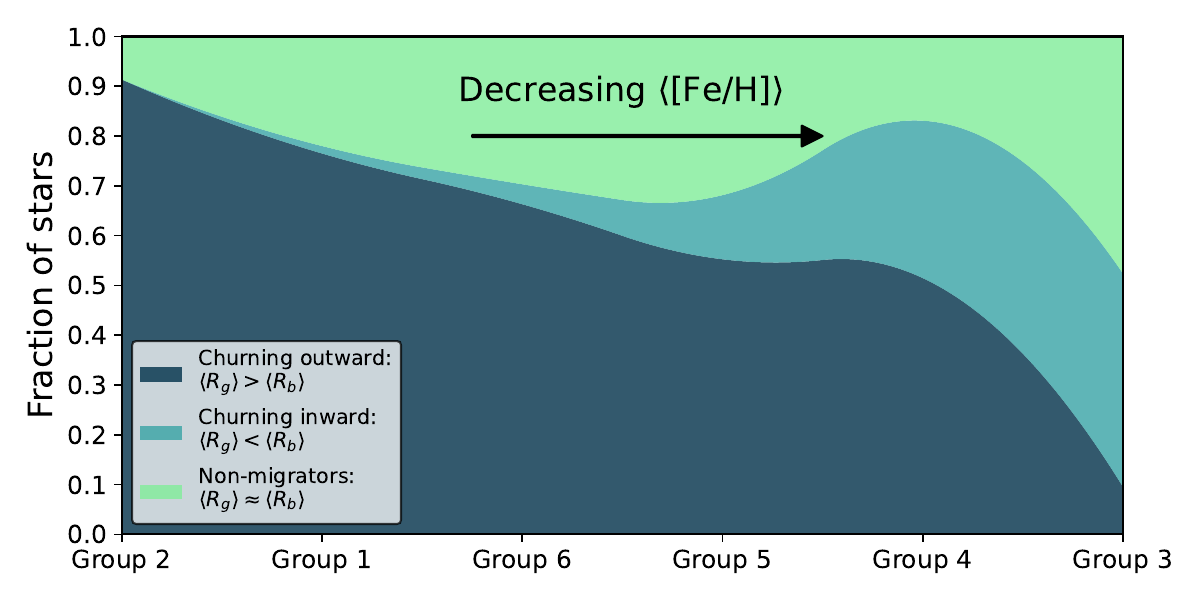}
    \caption{Fractional contribution of different migration behaviours across the chemically defined stellar groups. Ribbon plot showing the proportions of outward migrators (\rgui > \rbirth), inward migrators (\rgui < \rbirth), and stars whose guiding radii are consistent with their inferred birth radii (\rgui $\approx$ \rbirth). The groups are ordered from metal-rich (left) to more metal-poor (right). Outward migration dominates at super-metal-rich abundances, while the metal-poor groups display increasing fractions of inward migrators, illustrating how radial mixing varies systematically across the thin disc metallicity sequence.}
    \label{fig:ribbonplot}
\end{figure}

The chemically distinct groups identified in \citet{Dantas2023} provide a useful framework for examining how stars of different metallicities have contributed to the build-up of the Galactic thin disc. These groups span a wide range of abundance patterns and ages, reflecting the disc’s extended formation history and the diversity of enrichment pathways. By combining their chemical signatures with the \rbirth\ inferred from the GAM, we can evaluate how strongly each group has been affected by stellar mixing mechanisms and how these processes vary across the thin disc’s metallicity sequence.

The radial redistribution of these populations can be seen by comparing the distributions of \rbirth\ and \rgui\ directly, as shown in Fig. \ref{fig:boxenplot}. The side-by-side letter-value plots \citep[][]{Hofmann2017} show the median values and the spread of \rbirth\ and \rgui\ within each chemical group, ordered by decreasing \feh\ (from left to right). Since all stars have been recently observed in the Solar neighbourhood, their \rgui\ distributions are broadly similar to one another. However, in terms of \rbirth\, each group has a distinct behaviour. Metal-rich stars exhibit smaller \rbirth, being formed closer to the inner Galaxy, showcasing outward migration. As metallicity decreases, this trend slowly reverts, showing an increasingly more skewed distribution toward larger \rbirth. The metal-poor thin disc group displays the broadest birth-radius spread, consistent with a larger number of stars migrating inward. Altogether, the figure highlights the diversity of migration histories encoded in the chemically resolved thin disc.

The proportions of stars that have migrated outwards, inwards, or remained approximately at their birth radii are illustrated in Fig. \ref{fig:ribbonplot}. This visualisation highlights clear, coherent trends across the chemical groups. The super-metal-rich population (Group 2) is strongly dominated by outward migrators, consistent with their origin in the inner disc and subsequent displacement to larger guiding radii \citep[this particular Group was discussed in detail in][]{Dantas2023}. As metallicity decreases, the balance shifts: the intermediate groups show a higher fraction of stars whose \rgui\ are compatible with their inferred \rbirth, while the most metal-poor thin disc groups display increasing fractions of inward migrators. These contrasting trends reinforce the well-established link between metallicity and birth environment, while also showing how radial migration modulates this relationship across the thin disc.

Interpreting these patterns requires considering the interplay between chemical evolution, the inside-out growth of the disc, and the efficiency of angular-momentum redistribution. Metal-rich stars formed in the rapidly enriched inner regions are more susceptible to outward churning, which transports them into the Solar neighbourhood. Conversely, the more metal-poor groups contain older stars that likely formed over a broader radial range and at epochs when the radial metallicity gradient was steeper. Their present-day inward shifts are consistent with the combined action of churning and perhaps (mild) blurring, bringing stars from a range of formation radii into overlapping regions of the thin disc. Together, these trends align with the quantitative mixing estimates presented in \citet{Dantas2025a}, while Fig. \ref{fig:ribbonplot} complements them with a more intuitive view of the relative contributions of each migration class across metallicity.

In summary, the new visualisations presented here offer a complementary perspective on the findings of \citet{Dantas2025a}. By emphasising relative fractions and distribution shapes, rather than absolute counts or model-based diagnostics, these figures provide an accessible view of how different stellar populations have contributed to the modern thin disc. They reinforce the importance of chemically informed analyses and of flexible modelling tools such as GAMs in disentangling the disc’s complex chemo-dynamic assembly. Continued development of these approaches, together with larger and more precise spectroscopic samples, will further refine our understanding of the Galaxy’s long-term evolutionary history.

\begin{acknowledgements} 
MLLD acknowledges the support of Agencia Nacional de Investigación y Desarrollo (ANID), Chile, through Fondecyt Postdoctorado grant 3240344. MLLD and PBT also acknowledge ANID Basal Project FB210003. RS acknowledges funding from the National Science Centre, Poland, under project 2019/34/E/ST9/00133. RSS acknowledges the support from the São Paulo Research Foundation under project 2024/05315-4. PBT thanks Fondecyt Regular 2024/1240465.
\end{acknowledgements} 

\begin{refcontext}[sorting=nyt]
    \printbibliography
\end{refcontext}

\end{document}